\documentclass[aps,prb,reprint,superscriptaddress,
amsmath,amsfonts,amssymb,longbibliography]{revtex4-2}

\usepackage[utf8]{inputenc}
\usepackage[T1]{fontenc}

\usepackage{graphicx}
\usepackage{hyperref}
\usepackage{bm}
\usepackage{dcolumn}
\usepackage{multirow}

\usepackage{amsmath,amsfonts,mathtools}
\usepackage{physics}
\usepackage{empheq}
\usepackage{stackengine,scalerel,wasysym}

\usepackage[free-standing-units=true]{siunitx}

\usepackage[usenames,dvipsnames]{xcolor}
\usepackage[normalem]{ulem}
\usepackage{soul}
\usepackage{comment}

\makeatletter
\newsavebox\myboxA
\newsavebox\myboxB
\newlength\mylenA
\makeatother

\begin{document}

\title{  Connecting finite-size scaling and renormalization-group flows in Anderson localization on random graphs}

\author{Ignacio Garc\'ia-Mata}
\email{nacho.garcia.mata@gmail.com}
\affiliation{Instituto de Investigaciones F\'isicas de Mar del Plata (IFIMAR),
CONICET \& UNMdP, Funes 3350, B7602AYL Mar del Plata, Argentina.}
\affiliation{Consejo Nacional de Investigaciones Cient\'ificas y Tecnol\'ogicas (CONICET), Argentina.}
\author{Weitao Chen}
\affiliation{Department of Physics, University of Warwick, Coventry, CV4 7AL, United Kingdom.}
\author{John Martin}
\affiliation{Institut de Physique Nucl\'eaire, Atomique et de Spectroscopie,
CESAM, University of Li\`ege, B-4000 Li\`ege, Belgium.}
\author{Bertrand Georgeot}
\affiliation{Univ Toulouse, CNRS, Laboratoire de Physique Th\'eorique, Toulouse, France.}
\author{Gabriel Lemari\'e}%
\affiliation{INPHYNI, Universit\'e C\^ote d'Azur, CNRS, Nice, France}
\begin{abstract}
We investigate the relation between two complementary descriptions of the
Anderson transition on small-world random graphs: finite-size scaling of
eigenfunction moments and renormalization-group (RG) flows of multifractal
dimensions $D_q$.  The RG flows recover the qualitative structure previously
reported for the information dimension $D_1$, while their extension to other $D_q$
with smaller moment orders $q<1/2$ confirms the strongly multifractal nature of the localized phase characterized by a clear distinction between multifractal and localized scaling behaviors of different moments within the localized phase.
On the other hand, we show that we can construct another beta function characterizing the flow of eigenfunction moments, whose collapse onto a single function across different disorder strengths supports single-parameter scaling of the normalized moments within each sector.
These results raise an interesting question about the interpretation of two-parameter scaling: while a two-dimensional flow structure has been inferred from the information dimension \(D_1\) alone, our analysis reveals two distinct scaling sectors by considering different eigenfunction moments.
To characterize these properties, one should consider different observables, such as large $q>1/2$ and small $q<1/2$ eigenfunction moments, which are associated with distinct critical behaviors, in particular different critical exponents.
\end{abstract}
\maketitle
\section{Introduction}
\label{sec:intro}
Anderson localization on random graphs has attracted sustained attention
\cite{abouchacra,zirnbauer1986,Fyodorov1991,Mirlin1994,Monthus_2011,
biroli2012difference,deluca2014,Kravtsov_2015,Altshuler2016,
tikhonov2016Anderson,garciamata2017PRL,Tarquini2017,kravtsov2020localization, bandeira2026anderson},
motivated in part by its conceptual connection to many-body localization
\cite{Altshuler1997,BASKO20061126,Gornyi2005,TIKHONOV2021168525}.
Random graphs and trees provide a natural setting in which to test
whether the scaling ideas developed for finite-dimensional disordered
systems remain applicable when the effective dimensionality is infinite.
The graph-like structure of many-body Fock space makes single-particle
localization on random graphs a natural model for exploring related
questions of localization, ergodicity, and multifractality
\cite{Altshuler1997,BASKO20061126,Gornyi2005,TIKHONOV2021168525}.
Random graphs and trees have also long served as testing grounds for
the existence of a genuinely nonergodic delocalized phase.
In this context, the finite Cayley tree has been singled out for its
boundary-dominated physics
\cite{tikhonov2016fractality,sonner2017multifractality,KRAVTSOV2018148,
biroli2018delocalization}, in contrast to ensembles with loops and no
boundary, such as small-world, Erd\H{o}s--R\'enyi and random regular graphs, where the
delocalized phase is expected to be asymptotically ergodic
\cite{tikhonov2016Anderson,garciamata2017PRL,tikhonov2019statistics,
tikhonov2019critical,khaymovich2020fragile,sierant2023universality,
tarquini2016level}.
Despite their apparent simplicity, sparse random graphs combine broad
wave-function intensity distributions with logarithmic graph diameters,
leading to localization phenomena that differ qualitatively from those
in finite-dimensional lattices.
In particular, the separation between the graph volume and an effective
linear scale set by its logarithmically smaller diameter makes finite-size effects unusually
strong. Identifying the appropriate scaling variable and the asymptotic
regime is therefore a subtle task.

This difficulty bears on a long-standing question: how does the
conventional single-parameter scaling theory of Anderson
localization---the ``gang of four'' description of a flowing
dimensionless conductance \cite{gangoffour}, extended numerically to
finite-dimensional lattices
\cite{mackinnon1983scaling,asada2004numerical,PhysRevB.84.134209}---generalize
to graphs of effectively infinite dimension?
Two largely independent approaches have addressed this question.
The main purpose of this paper is to clarify their relationship,
particularly how each identifies and reveals the number of relevant
scaling parameters.

A finite-size scaling (FSS) description of the Anderson transition on
small-world and related random-graph ensembles has been developed using
eigenfunction moments and correlation functions
\cite{garciamata2017PRL,twocrit2020PRR,critical2022PRB}.
This approach exploits the fact that different moments probe different
regions of the eigenfunction intensity distribution.
Varying the moment order reveals the influence of rare branches in the
localized phase and critical regime, and distinguishes typical from
branch-averaged properties, which exhibit different critical behaviors
and exponents.
Once the critical finite-size dependence is divided out, these
observables display single-parameter scaling within the corresponding
regimes.
FSS thus provides access to the relevant scaling variables,
characteristic length and volume scales, and associated critical
exponents.
In particular, this approach has established that the localized phase
is governed by two distinct critical localization lengths:
one associated with propagation along rare branches and probed by
branch-averaged, large-$q$ moments; the other associated with the
typical bulk decay of the wave function transverse to these branches
and probed by small-$q$ moments.
However, a single observable is generically controlled by only one of
the two lengths at a time.
Revealing both therefore requires comparing observables of different
nature---branch-averaged and typical---corresponding to different
moment orders. 

An alternative, RG-inspired perspective has recently been proposed in
Refs.~\cite{vanoni2024pnas,altshuler2024renormal,niedda2024renormal}.
This approach is based on the finite-size flow of running fractal
dimensions, with particular emphasis on the information dimension $D_1$.
The associated $\beta$ function is defined as the logarithmic derivative
of $D_1$ with respect to system size, and the critical behavior is
analyzed through parametric trajectories in the $(D_1,\beta)$ plane.
Within this representation, the delocalized phase approaches an
effectively single-parameter flow, whereas the localized trajectories
form a size and disorder-dependent two-dimensional structure, interpreted as an
effective two-parameter flow.
This raises a natural question.
The two-length picture from FSS emerges from comparing different
observables, each individually compatible with single-parameter scaling
when expressed in terms of its appropriate scaling variable.
The RG approach, in contrast, reveals a two-dimensional flow structure
from a single observable, $D_1$, whose trajectories at different
disorder strengths do not collapse onto one curve.

Here we compare these two descriptions to explore the relationship between the two-dimensional structure of the \(D_1\)-flow trajectories and the two-length physics identified by FSS. In particular, we investigate how different observables reveal different aspects of the scaling behavior, and to what extent these descriptions can be reconciled.
We first examine how the FSS regimes depend on moment order and
construct the physical flows directly from the eigenfunction moments.
Near the information dimension, these flows recover the qualitative
structure reported in
Refs.~\cite{vanoni2024pnas,altshuler2024renormal}.
More generally, the flow topology depends on the moment order $q$ and
distinguishes the two localized regimes identified by FSS.
This already anticipates part of the answer: rather than a single,
observable-independent two-parameter flow, we find a moment-dependent
structure consistent with the distinction between branch-averaged
and typical behavior.

We then define a beta function for the normalized moments and
evaluate it directly from the unsmoothed data, without differentiating
a fitted scaling function.
The results for different disorder strengths collapse, to a very good
approximation, onto a common curve.
The behavior of this beta function near the fixed point retains
information about the critical exponents and distinguishes a
power-law divergence from the exponential divergence of the correlation
volume.
This normalized-moment beta function thus provides a direct
single-parameter description while retaining the critical information.

Finally, we explain how a two-dimensional flow structure can arise
from a single observable such as $D_1$, despite the single-parameter
scaling obeyed by the underlying moments: part of that structure
originates in the nontrivial size dependence of the critical behavior,
and part in the nonlinear relation between the natural FSS variable
and the system size, particularly in the critical and localized
regimes, where scaling is linear.
We make both contributions explicit in Appendix~\ref{sec:D1problem},
where we reconstruct the $D_1$ flow from the FSS scaling function
as a consistency check.

\section{Small-world random graph model}
\label{sec:model}
{We consider the Anderson tight-binding model on a small-world random graph (SWRG), constructed from a one-dimensional ring with nearest-neighbor hopping
and $\lfloor pN\rfloor$ additional long-range links. Here, $N$ is the number
of sites, $\lfloor\cdot\rfloor$ denotes the integer part, and
$p\in(0,1/2)$. The Hamiltonian reads
\cite{Chepelianskii,giraud2005quantum}
\begin{eqnarray}
\label{hamil}
H &=&
\sum_{i=1}^{N} \varepsilon_{i}|i\rangle\langle i|
+\sum_{\langle i,j\rangle}
\left(|i\rangle\langle j|+|j\rangle\langle i|\right)
\nonumber\\
&+&
\sum_{k=1}^{\lfloor pN\rfloor}
\left(|i_k\rangle\langle j_k|+|j_k\rangle\langle i_k|\right).
\end{eqnarray}
The onsite energies $\varepsilon_i$ are independent random variables drawn
from a Gaussian distribution with zero mean and standard deviation $W$,
unless otherwise stated. The second term describes the nearest-neighbor hopping along the ring, while the third introduces long-range hopping between randomly chosen pairs $(i_k,j_k)$ satisfying $|i_k-j_k|>1$.

For $p=0$, Eq.~\eqref{hamil} reduces to the standard one-dimensional Anderson model. For $p>0$, the long-range links give the graph a small-world structure.
On scales shorter than the typical distance between long-range links, $\ell_{\rm lr}\sim(2p)^{-1}$, the graph is locally quasi-one-dimensional.
At larger scales, repeated branching through the long-range links leads to an approximately exponential growth of the number of sites with graph distance.
This growth can be characterized by an effective branching number $K\simeq1+2p$  \cite{critical2022PRB,sierant2023universality}.

As a consequence, the graph diameter $d_N$ grows logarithmically with the volume,
\begin{equation}
d_N\sim\ln N,
\end{equation}
up to model-dependent prefactors. The system therefore possesses two natural finite-size variables: the total number of sites $N$, which defines a volume scale, and the graph diameter $d_N\sim\ln N$, which plays the role of an
effective linear size. This distinction is central to the finite-size scaling analysis below, where different observables and different regions of the
phase diagram may be governed either by volumic or by linear scaling.

A crucial difference between the SWRG and the Cayley tree or Bethe-lattice geometry is that the SWRG has no boundary and contains loops, whose typical
length is the finite graph diameter. This distinction is important for the nature of the delocalized phase. On trees, boundary effects have been argued to support a nonergodic delocalized regime
\cite{tikhonov2016fractality,sonner2017multifractality,KRAVTSOV2018148,biroli2018delocalization}.
By contrast, in random graphs with loops and no boundaries, including SWRGs and
Erd\H{o}s--R\'enyi-type ensembles, the available evidence supports an
asymptotically ergodic delocalized phase, with multifractal behavior restricted
to finite sizes and/or to the critical region
\cite{tikhonov2016Anderson,garciamata2017PRL,
tikhonov2019statistics,tikhonov2019critical,
khaymovich2020fragile,sierant2023universality,
garciagarcia2007dimensional,tarquini2016level}.

Throughout this work, the parameter $p$ controls  the effective local connectivity, while preserving both the
small-world character of the graph and the logarithmic relation between its
diameter and volume.}
\section{Scaling of eigenfunction moments}
\label{sec:moments}

Throughout, we denote by $N$ the volume, or number of sites, and use
\begin{equation}
L\equiv \ln N
\end{equation}
as the running scale. For the random graphs considered here, $L$ is
proportional, up to model-dependent constants, to the graph diameter,
$d_N\sim \ln N$.

The Anderson transition can be efficiently characterized
through the finite-size scaling of eigenfunction moments, see e.g. \cite{RevModPhys.80.1355, PhysRevB.84.134209} in finite dimension and \cite{garciamata2017PRL,twocrit2020PRR,critical2022PRB} for small-world graphs.
For a normalized eigenstate $\ket{\psi}$, we define the disorder-averaged
$q$th moment
\begin{equation}
\label{eq:moments}
P_q(N,W)=\left\langle \sum_i |\psi_i|^{2q} \right\rangle ,
\end{equation}
where the average is taken over disorder realizations and eigenstates in a narrow energy window around the band center. Different values of $q$ probe
different regions of the eigenfunction intensity distribution: large $q$
emphasizes the largest amplitudes, whereas small $q$ gives increasing weight
to the small-amplitude components.

In the limit of large $N$, whenever a multifractal behavior is
reached, one has
\begin{equation}
P_q \sim N^{-\tau_q}, \qquad D_q=\frac{\tau_q}{q-1},
\end{equation}
where $D_q$ denotes the generalized multifractal dimension. For finite sizes,
it is convenient to introduce the running, or system-size dependent, exponents
\begin{eqnarray}
\label{eq:running_tauD}
\tau_q(N,W)&\equiv&
-\frac{d\ln P_q}{d\ln N}
=-\frac{d\ln P_q}{dL},
\\
\label{eq:running_D}
D_q(N,W)&\equiv&
\frac{\tau_q(N,W)}{q-1}
=-\frac{1}{q-1}\frac{d\ln P_q}{d\ln N},
\end{eqnarray}
which approach their asymptotic values as $N\to\infty$.


The SWRG system undergoes a delocalization-localization transition at a critical disorder $W_c$, which depends on the long-range-link parameter $p$ \cite{Chepelianskii,giraud2005quantum,garciamata2017PRL}.
For weak disorder, $W<W_c$, the system is delocalized and asymptotically ergodic, with $\tau_q\to q-1$ and $D_q\to1$. For strong disorder, $W>W_c$, the system is localized and $D_q\to0$ for sufficiently large $q$. However, the approach to these limits is highly nontrivial because of the infinite effective dimensionality of the graph and the logarithmic relation
between its diameter and volume \cite{PhysRevResearch.6.L032024,PhysRevB.110.014210}.

In the localized regime of SWRGs, the localized eigenstates are strongly
anisotropic: rare branches can support large amplitudes over a much larger distance
than typical directions in the graph. This leads to two characteristic
localization lengths, associated with propagation along the rare branches $\xi_\parallel$, or
transverse to them $\xi_\perp$, which manifest themselves differently in typical or
branch (instead of the commonly disorder)-averaged observables \cite{critical2022PRB}. This can be related to the fact that asymptotic multifractal exponents $D_q(N,W)$, Eq.~\eqref{eq:running_D}, display strong
multifractality,
\begin{equation}\label{eqstrongMF}
\tau_q(W)\simeq
\begin{cases}
q/q^*(W)-1, & q \le q^*(W),\\
0, & q \ge q^*(W),
\end{cases}
\end{equation}
with $q^*(W)\le 1/2$. 
For $q<q^*(W)$, $P_q$ moments are dominated by the small amplitudes of the eigenfunctions and have a non-zero multifractal dimension, whereas for $q>q^*(W)$ $P_q$ shows characteristics of a localized behavior with $D_q \rightarrow 0$. 
We emphasize that these different scaling behaviors occur within the same localized phase and do not correspond to distinct thermodynamic phases.
$q^*(W)$ is related to the typical localization length,
$ q^*(W)\simeq \xi_\perp \ln K$ \cite{critical2022PRB}, see also \cite{Mirlin1994}.
Analogous characteristic
times may be defined for dynamical probes \cite{chen2026critical}.

At the critical point,
\begin{equation}
q_c^*\equiv q^*(W_c)=\frac{1}{2},
\end{equation}
\cite{PhysRevResearch.6.L032024,PhysRevB.110.014210},
consistently with the multifractal symmetry
$\Delta_q=\Delta_{1-q}$ \cite{mirlin2006Exact,bilen2021symmetry}. This means that $D_q>0$ for $q<q^*_c=1/2$, a standard multifractal critical behavior, whereas for $q>q^*_c$, $D_q \rightarrow 0$. It turns out that the system size dependence of $P_q^c(N)$ for $q>1/2$ in random graphs of infinite effective dimension is highly non-trivial, in particular not algebraic in $N$ contrary to the case of finite dimension, see \cite{Mirlin1994, TIKHONOV2021168525, PhysRevResearch.6.L032024, PhysRevB.110.014210}. Two types of critical behaviors have been predicted and observed numerically: 
\begin{equation}\label{eqcritPq}
P_q(N)\simeq
\begin{cases}
(\ln N)^{-d_q (q-1)}\\
P_q(\infty) + c (\ln N)^{-\mu} \; .
\end{cases}
\end{equation}
The first case has been called logarithmic multifractality, characterized by logarithmic multifractal dimensions $d_q$. This is the critical behavior observed in SWRGs, and we will focus on this behavior in the following. The second case, called critical localization, has been observed for random regular graphs with an exponent $\mu=1/2$.

Finally, in the delocalized phase,
$q^*$ approaches its ergodic value $q^*=1$ at sufficiently large system sizes
\cite{critical2022PRB}.

The central object of the finite-size scaling analysis is the normalized
moment
\begin{equation}
\label{eq:Fq_def}
F_q(N,W)=\frac{P_q(N,W)}{P_q^c(N)},
\qquad
P_q^c(N)\equiv P_q(N,W_c),
\end{equation}
which divides out the critical finite-size dependence. In finite dimensions, this normalization by the critical behavior $P_q^c$ is crucial to obtain single parameter scaling, see e.g. \cite{PhysRevB.84.134209}. The finite-size scaling assumption for random graphs generalizes this scaling assumption. 

Because $d_N\sim\ln N$, scaling as a function of the ratios $L/\xi\propto d_N/\xi$ or $N/\Lambda$ are not equivalent. Hence, these two distinct scaling variables arise naturally,
\begin{equation}
\label{eq:Fdeq}
\frac{P_q(N,W)}{P_q^c(N)}
=
F_q\!\left(\frac{d_N}{\xi(W)},\frac{N}{\Lambda(W)}\right),
\end{equation}
corresponding to linear and volumic scaling, respectively. Crucially, depending on $q$,
the disorder strength, and the side of the transition, one of these variables
provides the dominant finite-size dependence. 
For large moment orders, and in particular for $q>1/2$, the delocalized
phase is governed by volumic scaling, with a correlation volume
\begin{equation}\label{eq:Lambdadiv}
\Lambda(W)\sim
\exp\!\left[\alpha(W_c-W)^{-\nu_{\rm deloc}}\right],
\qquad
\nu_{\rm deloc}\simeq \frac12 .
\end{equation}
On the localized side, the dominant scaling is instead linear, with a
localization length
\begin{equation}
\xi(W)=\xi_\parallel\sim (W-W_c)^{-\nu_{\rm loc}},
\qquad
\nu_{\rm loc}\simeq 1 .
\end{equation}
For small $q<1/2$, the small-amplitude elements lead to a different scaling
structure. In the accessible finite-size regime the data are organized by a linear scaling variable on both sides of the transition, with exponents $\nu_{\rm loc/deloc}\approx 1/2$. One can show that this implies the following behavior for the typical localization length $\xi_\perp$:
\begin{equation}
    \xi_\perp^{-1} = {\xi_\perp^c}^{-1} + c (W-W_c)^{1/2} \;, 
\end{equation}
for $W>W_c$, with $\xi_\perp^c = 1/(2 \ln K)$ a universal number, and the square-root singularity is related to the critical exponent $1/2$.

\begin{figure}[t]
    \centering
    \includegraphics[width=0.95\linewidth]{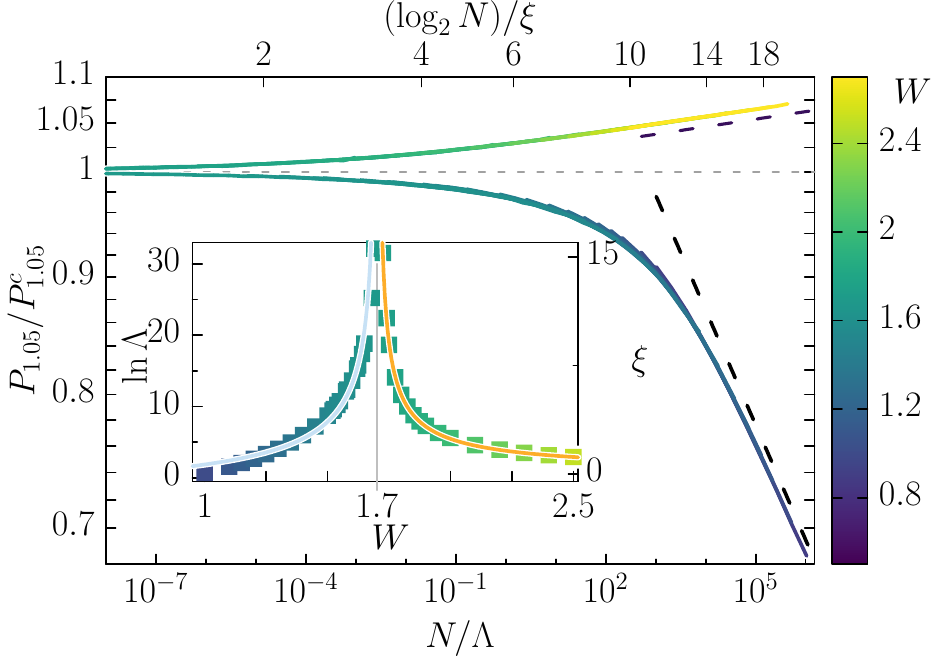}
    \caption{Finite-size scaling of $F_q=P_q/P_q^c$  (log-scale) for the SWRG for $p=0.06$, $W_c\approx 1.7$ and   $q=1.05$. The system sizes used for the scaling are $N=2^9,\ldots 2^{20}$. The delocalized branch is collapsed using the volumic variable $N/\Lambda$, i.e. the ratio of the system volume with the correlation volume, while the localized branch is collapsed using the linear variable $d_N/\xi$, i.e. the ratio the linear system size, $\sim \log_2 N$, over the localization length $\xi$. Dashed lines indicate the expected asymptotic behaviors, in particular the ergodic behavior in the lower delocalized branch. The inset shows $\ln\Lambda(W)$ and $\xi(W)$. The delocalized data are consistent with $\ln\Lambda\sim(W_c-W)^{-1/2}$. The localized fit gives an  exponent $\nu_{\rm loc}\simeq0.74$. We attribute the difference between the fitted critical exponent $\nu_{\rm loc}\simeq0.74$ and the expected value $1$ to the proximity of $q=1.05$ to $q=1/2$ where contributions from branch-averaged and typical localization are not fully separated.
    \label{fig-scaling105}}
\end{figure}

Below we show two numerical examples of the different scalings for two different $q$ values. The beta function approach of Ref.~\cite{vanoni2024pnas} relies on the  information dimension which corresponds to the limit $q\to1$. Since $P_1=1$ by normalization, we use $q=1.05$ which is close enough to $1$. Then we also show results for a small $q$ case.

In Fig.~\ref{fig-scaling105} we show how $P_q/P_q^c$  collapses to a single parameter scaling function $F_q$ (as expressed in Eq.~(\ref{eq:Fdeq})), for  
 $q=1.05$ and $p=0.06$. As shown for $q=2$ in Refs.~\cite{garciamata2017PRL,critical2022PRB} the delocalized phase (lower branch, and lower $x$-axis) is better collapsed into one curve by a volumic scaling. In the top branch of the main panel we show the localized phase is compatible with 
a linear scaling. In the inset we show that the correlation volume is well approximated with a critical exponent (fixed) $\nu_{\rm deloc}=1/2$. The asymptotic ergodic limit $F_q\sim N^{-(q-1)}$ can be clearly seen through the (eye-guiding) dashed line. In the localized phase we get $\nu_{\rm loc}\approx 0.74$. The numerical deviation from the expected  $\nu_{\rm loc}\approx 1$ can arise from the fact that $q$ is not large enough and the scaling is not fully linear but a combination of both linear and volumic. 

\begin{figure}[t]
    \centering
    \includegraphics[width=0.95\linewidth]{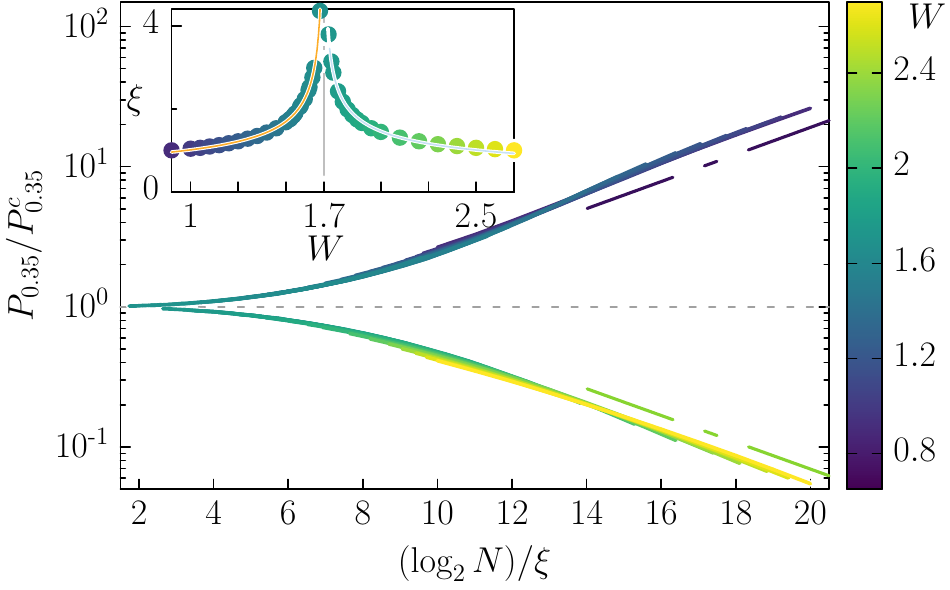}
  \caption{Finite-size scaling of $F_q=P_q/P_q^c$ for $p=0.06$, $W_c\approx 1.7$ and  $q=0.35$. The system sizes used are the same as in Fig.~\ref{fig-scaling105}.  
Both sides of the transition are collapsed using the effective linear
scaling variable. Dashed lines indicate the expected asymptotic slopes.
The inset shows the fitted length scale $\xi(W)$, where we find the exponents
$\nu_{\rm loc/deloc}\approx 0.4$ (close to 0.5).
    \label{fig-scalingsmallq}}
\end{figure}

We now discuss the scaling for $q=0.35<1/2$ shown in  Fig.~\ref{fig-scalingsmallq}. In this case both phases are well described by a linear scaling. In the inset we show that the numerical value we get for the critical exponent is $\nu_{\rm loc/deloc}\approx 0.4$, which is close to  the expected $1/2$. There are however subtleties concerning the localized phase that need to be discussed, due to their relevance further in this work. The localized regime is characterized by a strong multifractal behavior, see Eq.~\eqref{eqstrongMF}, with different values of $W$ having different $q^*(W)$ values. We can thus define a disorder bound $W^*$, such that $q^*(W^*)=q$ dividing asymptotic (i.e. when $N\rightarrow \infty$) multifractal $D_q>0 $ or localized $D_q=0$ behaviors in the localized phase. The linear scaling shown in Fig.~\ref{fig-scalingsmallq} describes the finite-size scaling close to $W_c$, for $W<W^* \approx 2$ for $q=0.35$. For $W>W^*(q=0.35)\approx 2$, volumic scaling instead describes better the data and is responsible for an asymptotic $D_q=0$ localized behavior.

In the following sections, we use
these single-parameter scaling curves as the starting point for the
$\beta$-function analysis. We first recover, for large moments, the flow
previously obtained from the information dimension and then extend the
construction to small $q$, where the $\beta_q$ trajectories reproduce the richer structure of the localized phase and its separation at $W^*(q)$.
\section{Beta-function approach}
\label{sec:beta}
In Ref.~\cite{vanoni2024pnas}, an RG description of the Anderson transition
on random regular graphs was formulated in terms of two running quantities.
Under a real-space blocking transformation, a subtree of depth $L$ becomes an
effective site whose connectivity, denoted here by $\mathcal K$ to distinguish
it from the bare branching number {$K_0=K$}, grows as $\mathcal K(L)=K_0^L$.  The second
quantity is the running information dimension
\begin{equation}
D_1(\mathcal K,W)=\frac{dS_1}{d\ln\mathcal K},
\end{equation}
obtained from the finite-size Shannon entropy{:
\begin{equation}
    S_1 (N,W)= - \langle \sum_i \vert \psi_i\vert^2 \ln \vert \psi_i\vert^2 \rangle \; .
\end{equation}}
The associated flow is
described by
\begin{equation}
\beta_D(D_1,\mathcal K)=
\frac{d\ln D_1}{d\ln\mathcal K}.
\end{equation}
{This approach aims at generalizing the famous scaling theory of the ``gang of four'' \cite{gangoffour}, considering $D_1$ as the equivalent of the dimensionless conductance.}

{A key observation of \cite{vanoni2024pnas} is that the obtained trajectories in the $(D_1,\beta_D)$ plane do not collapse onto a single scaling curve, which appears inconsistent with single-parameter scaling that we will discuss in the following. }
In turn, Ref.~\cite{vanoni2024pnas} decomposes the
flow as
\begin{equation}
\beta_D(D_1,\mathcal K)=\beta_0(D_1)
+\beta_{\rm fs}(D_1,\mathcal K),
\end{equation}
where $\beta_0$ is the size-independent envelope governing the asymptotic
{single}-parameter flow on the delocalized side, while $\beta_{\rm fs}\leq0$
contains the explicit finite-size dependence (this term is denoted $\beta_1$
in Ref.~\cite{vanoni2024pnas}).  The ergodic fixed point is located at
$(D_1=1, \beta_D=0)$.  In the localized phase they propose
$D_1\sim\mathcal K^{-\alpha(W)}$, so that $D_1=0$ forms a line of localized
fixed points with $\beta_D\to-\alpha(W)$; the transition corresponds to the
terminal point $\alpha(W_c)=0$.

This construction is used in Ref.~\cite{vanoni2024pnas} {to understand} finite-size effects {across the Anderson transition on random regular graphs}.  In the delocalized phase the proposed
envelope satisfies $\beta_0(D_1)>0$ for $0<D_1<1$, so every sufficiently large
RRG ultimately flows to the ergodic fixed point.  The multifractal behavior
observed at finite size is therefore interpreted as a crossover rather than
as {a signature of} a stable nonergodic extended phase.  Close to the transition, the negative
finite-size contribution $\beta_{\rm fs}$ can initially dominate and drive
$D_1$ downward; as the running connectivity increases, this contribution
weakens, $D_1$ passes through a minimum, and the positive single-parameter term
$\beta_0$ subsequently drives the flow toward $D_1=1$.  The competition
between the two terms thus accounts for the nonmonotonic size dependence of
eigenstate and spectral observables near the transition {(see also \cite{tikhonov2016Anderson, PhysRevB.105.094202})}.

On the critical trajectory Ref.~\cite{vanoni2024pnas} obtains
$D_1^c\sim(\ln\mathcal K)^{-2}$ and interprets it as a marginally irrelevant
flow.  The behavior of $\beta_0(D_1)$ close to $D_1=0$ is not fixed by the
available numerics, and two scenarios are considered.  If
$\beta_0(D_1)\propto D_1$, an additional large scale with exponent $\nu=1$
appears after the scale with $\nu=1/2$ associated with reaching the
single-parameter trajectory.  If instead
$\beta_0(D_1)\propto\sqrt{D_1}$, no additional scale arises and the only
exponent is $\nu=1/2$.  This structure, together with the localized line of
fixed points terminating at criticality, motivates their analogy with a
Kosterlitz--Thouless flow.  Ref.~\cite{vanoni2024pnas} stresses that present
data do not distinguish the two small-$D_1$ scenarios.

{Our construction is equivalent in spirit but uses quantities naturally
available from the eigenfunction moments.  At the scale of the full graph,
$\mathcal K\sim K_0^{d_N}\sim N$, and hence $\ln\mathcal K$ differs from
$\ln N$ only by an inessential scale convention.  We replace the entropy
derivative by the running multifractal dimension
\begin{equation}
\label{eq:Dq}
D_q(N,W)=-\frac{1}{q-1}\frac{d\ln P_q(N,W)}{d\ln N}.
\end{equation}
This reduces to the information dimension as $q\to1$ but also allows us to
resolve the moment-dependent structure of the localized phase.  Writing
$x=\log_2N$, we define
\begin{equation}
\label{eq:beta}
\beta_q(N,W)=\frac{d\ln D_q}{d\ln N}
=\frac{1}{\ln2}\frac{d\ln D_q}{dx}.
\end{equation}
The flow diagram is obtained by eliminating $N$ between $D_q(N,W)$ and
$\beta_q(N,W)$ at fixed $W$.  Thus our $(D_q,\beta_q)$ trajectories are the
moment-resolved counterparts of the $(D_1,\beta_D)$ flow of
Ref.~\cite{vanoni2024pnas}, with $N$ playing the role of the running
connectivity at the scale of the complete graph.  Related RG formulations
are discussed in Refs.~\cite{altshuler2024renormal,niedda2024renormal}.  We
adopt this flow representation, but do not assume the same critical or
localized asymptotic functions: those are fixed below using the moment
scaling previously established for the SWRG model.}

{The numerical extraction of this flow requires some care.  Equation
\eqref{eq:Dq} {for the $N$-dependent multifractal dimension $D_q(N,W)$} is already a logarithmic derivative of the measured moments;
evaluating Eq.~\eqref{eq:beta} {for $\beta_q(N,W)$} directly would amount to taking a second
numerical derivative.  This procedure strongly amplifies statistical
fluctuations, particularly in the localized phase, where $P_q$ approaches a
constant and $D_q$ becomes small.  We therefore compute $D_q$ by finite
differences between consecutive sizes, apply two successive moving averages
with a two-point window, fit the resulting $D_q(x,W)$ curves, and evaluate
$\beta_q$ analytically from the fitted functions.  The smoothing precedes all
fits and is used only in this beta-function analysis; the finite-size scaling
of the moments presented above does not rely on it.

We could in principle propagate the uncertainties from the sample variance of $P_q$ to  the unsmoothed finite-difference. However,  the moving average correlates neighboring
points.  This is
important deep in the localized phase: there, $D_q$ is obtained by
subtracting two nearly equal values of $\ln P_q$, while the number of disorder
realizations decreases with increasing $N$.  The largest-size derivatives
can therefore fluctuate visibly even when the underlying $P_q$ data already
show a stable asymptotic trend.

We make two types of fits{: we} choose 
 to reproduce the asymptotic behavior of $P_q$ established in our
previous analyses \cite{garciamata2017PRL,twocrit2020PRR,critical2022PRB}, while allowing the leading finite-size corrections needed
over the accessible range.  In particular, Pad\'e functions are used only on
the delocalized side (here we follow Ref.~\cite{vanoni2024pnas}), where the limiting value $D_q=1$ is known.  On the
localized side we instead use explicit correction expansions with the
appropriate thermodynamic limit.  This avoids spurious behavior, like $D_q<0$ at large $N$.}
\subsection{Information-dimension flow}

\label{sec:beta_largeq}

Since $P_1=1$, Eq.~\eqref{eq:Dq} cannot be evaluated directly at $q=1$, we first form
the finite-difference dimensions at the two closest available moments  (we have data for $q\in [-4,4]$ every $\delta q=0.05$) and
then take
\begin{equation}
\label{eq:D1_symmetric}
D_1(N,W)\simeq\frac{D_{0.95}(N,W)+D_{1.05}(N,W)}{2}.
\end{equation}
The order of these operations matters: $D_{0.95}$ and $D_{1.05}$ are computed
separately from their respective $P_q$ data, Eq.~\eqref{eq:D1_symmetric} is
then evaluated at their common size midpoints, and only the resulting $D_1$
curve is smoothed and fitted.

{For $W<W_c$, following \cite{vanoni2024pnas}, we fit $D_1(x,W)$ ($x\equiv \log_2 N$) with a second-order Pad\'e approximant
constrained to approach unity as $x\to\infty$.  
 {At the critical point, analytical and numerical results \cite{garciamata2017PRL,critical2022PRB, PhysRevResearch.6.L032024, PhysRevB.110.014210} support a logarithmic multifractal behavior for $q>1/2$:
\begin{equation}
\label{eq:Pq_critical_log}
P_q^c(N)\simeq B_q(\ln N)^{-\sigma_q},
\end{equation}
with $\sigma_q = d_q(q-1)$, $d_q$ the logarithmic multifractal dimension, see Eq.~\eqref{eqcritPq}.}
Substitution into Eq.~\eqref{eq:Dq} yields
\begin{equation}
\label{eq:Dq_critical_log}
D_q^c(N)\simeq
\frac{\sigma_q}{q-1}\frac{1}{\ln N}.
\end{equation}
In the limit $q\to1$, the ratio
$s_1\equiv\lim_{q\to1}\sigma_q/(q-1)$ remains finite, so that
\begin{equation}
\label{eq:D1_critical_fit}
D_1^c(N)\simeq\frac{s_1}{\ln N},
\qquad
\beta_1^c(N)\simeq-\frac{1}{\ln N}\longrightarrow0.
\end{equation}
We implement the $q\to1$ limit as expressed in 
 Eq.~\eqref{eq:D1_symmetric}.  Eliminating $\ln N$ between
the two expressions in Eq.~\eqref{eq:D1_critical_fit} gives
\begin{equation}
\label{eq:beta1_critical_line}
\beta_1^c(D_1)\simeq-\frac{D_1}{s_1}.
\end{equation}
The critical trajectory is therefore asymptotically a straight line through
the origin in the $(D_1,\beta_1)$ plane.  Subleading inverse-logarithmic
corrections can generate a weak curvature over the accessible sizes, but do
not change this limiting linear behavior.

In the localized phase, {we rely on the strongly anisotropic
localization picture} established in Ref.~\cite{critical2022PRB}.  Moments above $q=1$ probe  large amplitudes of the eigenfunctions, i.e. the rare
populated branches, along which the wave function is exponentially localized
with FSS length {$\xi(W)=\xi_\parallel\sim (W-W_c)^{-1}$} which diverges at the transition with a critical exponent $1$.  A finite graph truncates such a
branch at a distance of order $d_N\simeq\ln N/\ln K$.  The leading missing-tail
correction to a localized moment is therefore expected to behave as
\begin{align}
P_q(N,W)-P_q^\infty(W)
&\propto \exp\left[-\frac{\kappa_q d_N}{\xi(W)}\right]
\nonumber\\
&\sim N^{-\omega_q(W)},
\label{eq:Pq_rare_branch_correction}
\end{align}
where
\begin{equation}
\label{eq:omega_xi}
\omega_q(W)\simeq
\frac{\kappa_q}{\xi(W)\ln K}.
\end{equation}
Here $\kappa_q$ is a moment-dependent coefficient describing the decay of the
corresponding tail. Equation~\eqref{eq:omega_xi} is not imposed in
the fit, but provides a direct physical motivation for a power of $N$ and
connects its exponent with the inverse FSS length.

Differentiating Eq.~\eqref{eq:Pq_rare_branch_correction} shows that the
running dimension has the same leading power of $N$.  For the symmetrized
information dimension we consequently fit
\begin{equation}
\label{eq:D1_localized_fit}
D_1(N,W)=C(W)N^{-\omega(W)}
\left[1+a(W)N^{-\omega(W)}\right].
\end{equation}
The second factor in Eq.~\eqref{eq:D1_localized_fit} retains the first
subleading power generated by expanding about the localized plateau.  It
preserves $D_1>0$ over the fitted interval and the asymptotic limit
$D_1\to0$.  Differentiating the fitted function analytically gives
\begin{equation}
\label{eq:beta1_localized_fit}
\beta_1(N,W)=-\omega-
\frac{\omega aN^{-\omega}}{1+aN^{-\omega}},
\end{equation}
so that
\begin{equation}
\label{eq:beta1_localized_limit}
D_1\longrightarrow0,
\qquad
\beta_1\longrightarrow-\omega(W)
\end{equation}
in the thermodynamic limit.
Since $\xi$ diverges on approaching $W_c$, the interpretation in
Eq.~\eqref{eq:omega_xi} implies $\omega\to0$, continuously connecting
the localized limit $\beta_1\to-\omega$ to the critical behavior
$\beta_1^c\to0$ derived from Eq.~\eqref{eq:Pq_critical_log}.}

{Figure~\ref{fig:D1_fits} shows the smoothed dimensions $D_1(N,W)$ (circles) together with the corresponding fitting functions.  The fit curves capture the systematic size dependence;
deviations of isolated large-$N$ points in the far-localized regime reflect
the rapidly increasing relative uncertainty of the numerical derivative. It is important to remark that for larger system sizes the number of samples averaged is smaller, leading to larger uncertainty.

\begin{figure}[t]
    \centering
    \includegraphics[width=0.95\linewidth]{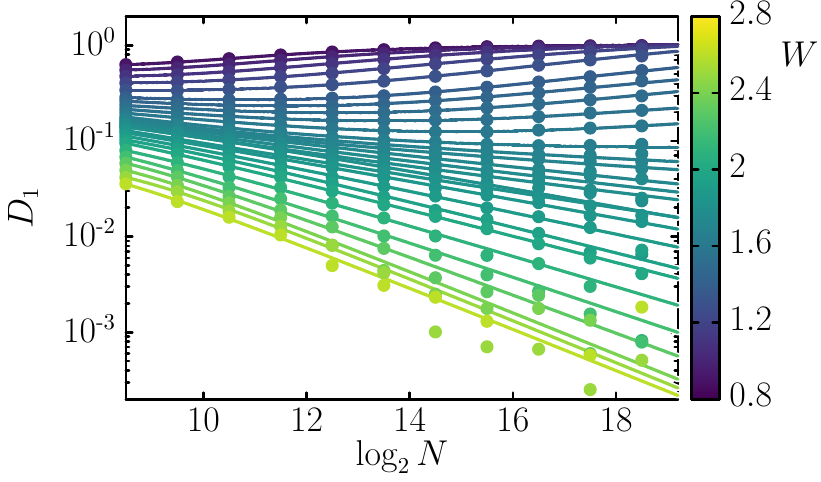}
    \caption{Smoothed running {multifractal} information dimension $D_1$ as a function of
    $x=\log_2N$, for $p=0.06$.  The symbols are obtained from
    $D_1=(D_{0.95}+D_{1.05})/2$ after the two finite differences have been
    formed; the solid curves are the {localized/delocalized} phase-dependent fits described in the text, Eq.~\eqref{eq:D1_localized_fit} in the localized phase, Eq.~\eqref{eq:Dq_critical_log} at criticality and second-order Padé approximant in the delocalized phase.
    Only the indicated system-size interval is included
    in the fits. 
    }
    \label{fig:D1_fits}
\end{figure}

\begin{figure}[t]
    \centering
    \includegraphics[width=0.95\linewidth]{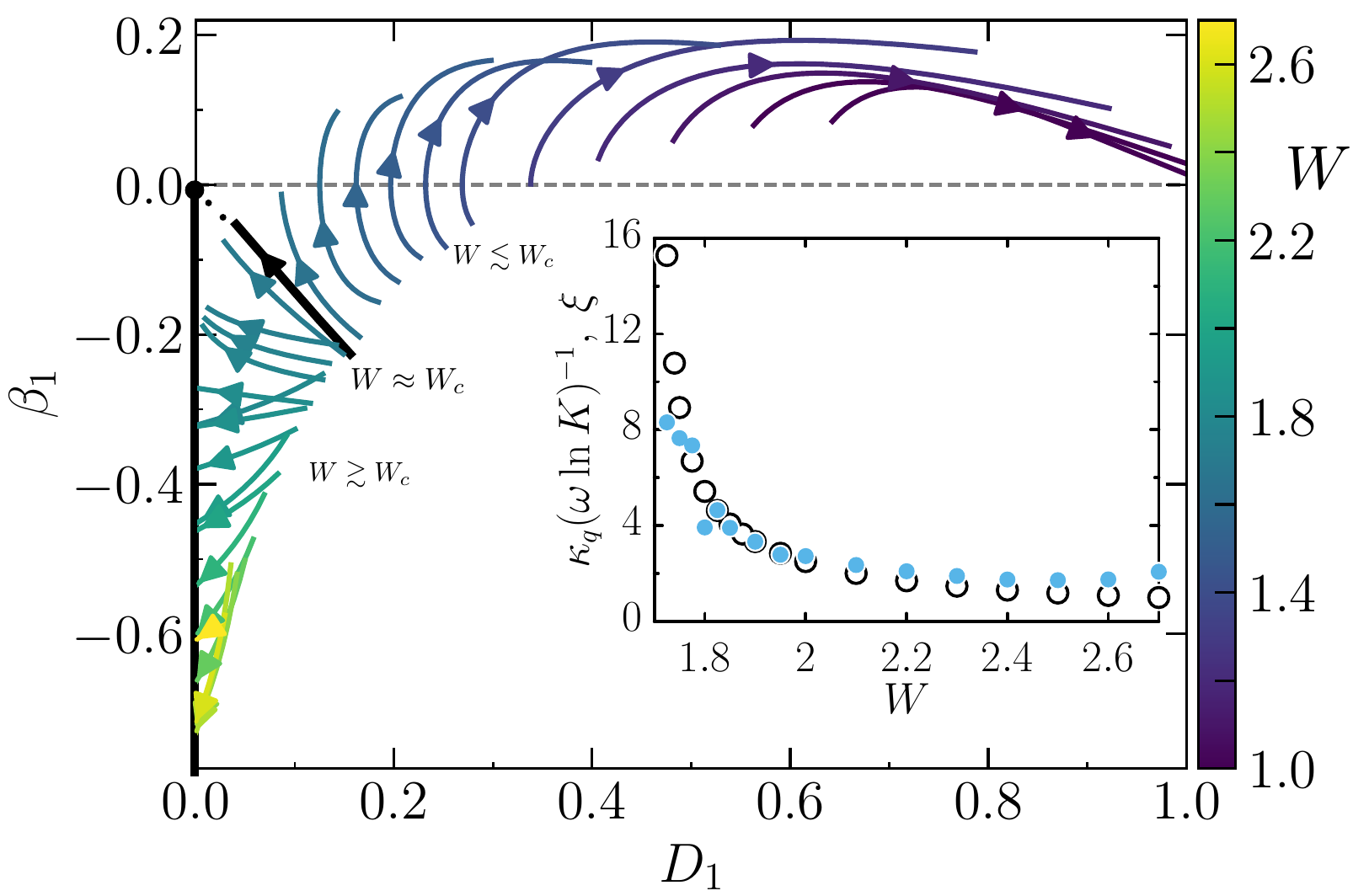}
    \caption{Flow diagram $\beta_1(D_1)$ obtained by differentiating the
    fitted smooth curves of Fig.~\ref{fig:D1_fits} analytically. The beta function
    is not computed by a second finite difference of the data because it would be too sensitive to fluctuations . Each trajectory
    corresponds to a fixed disorder and increasing {system} size. Inset: $\kappa_q(\omega(W)\ln K)^{-1}$ obtained from the limit $D_q\to 0$ of $\beta_1$ compared to  the localization length $\xi$ obtained from FSS with $q=1.05$ and $W_c=1.7${, see Fig.~\ref{fig-scaling105}. $K$ is the connectivity $K\approx1+2p \approx 1.12$ for $p=0.06$ \cite{critical2022PRB}.}
    \label{fig-beta}}
\end{figure}

The corresponding flow is shown in Fig.~\ref{fig-beta}.  
In the localized phase, the trajectories flow toward \(D_q\to0\), whereas in the delocalized phase they ultimately approach the ergodic fixed point \(D_q=1\).
At  criticality, as expressed in Eq.~(\ref{eq:beta1_critical_line}), the flow is almost a straight line (thick black line) that goes to 0. The power-law  fit for the localized phase makes explicit
the finite limiting value $\beta_1\to-\omega(W)$ suggested by exponential
localization on the graph. In the inset of Fig.~\ref{fig-beta} we show the limiting value $\omega(W)$ obtained from fitting the smoothed data for $D_1$ with the function of Eq.~(\ref{eq:D1_localized_fit}), and compare it with the FSS length $\xi(W)$ for $q=1.05$, through Eq.~\eqref{eq:omega_xi}.

In Ref.~\cite{vanoni2024pnas}, the running dimension in the localized phase
is -- like we do here -- proposed to decay algebraically with the running connectivity,
$D(K)\sim K^{-\alpha(W)}$, which implies a disorder-dependent limiting value
$\beta\to-\alpha(W)$ as $D\to0$.  This asymptotic behavior is not directly
resolved by the numerical flow reported there: the localized trajectories
appear compatible with an approach $\beta\to0^{-}$, although the authors
emphasize that the small-$D$ region depends on the interpolation procedure.
In particular, their numerical beta function is obtained by fitting the
available $D(K)$ data with a fourth-order polynomial in $1/\ln K$, used only
within the accessible size interval.  If extrapolated, any finite polynomial
of this form would give $\beta\to0$ and therefore cannot establish a finite
negative intercept.  The distinction between $\beta\to0^{-}$ and
$\beta\to-\alpha(W)$ consequently remains an asymptotic question beyond the
sizes studied in Ref.~\cite{vanoni2024pnas}.  

{Here we have shown that the latter scenario, namely the power-law decay of $D_1$ with $N$ in Eq.~\eqref{eq:D1_localized_fit}, motivated independently by exponential localization along rare branches \cite{twocrit2020PRR,critical2022PRB}, provides a good description of our numerical data. We note also that the critical behavior $\beta_1^c(D_1) \propto D_1$ in Eq.~\eqref{eq:beta1_critical_line} that we observe is compatible, within the description of Ref.~\cite{vanoni2024pnas}, with a localization-length critical exponent of $1$. This is again consistent with our observation that $\omega_q(W) \propto \xi(W)^{-1} = \xi_\parallel^{-1}$, as expressed in Eq.~\eqref{eq:omega_xi} and clearly demonstrated in the inset of Fig.~\ref{fig-beta}.
 }

An important claim of Ref.~\cite{vanoni2024pnas} is that the fact that the different curves of the RG flow in the $(\beta_1,D_1)$ plane do not collapse onto each other is a signature, at least in the localized phase, of a two-parameter scaling flow. We note, however, that all the data we have used to produce such a flow do follow a single-parameter scaling function, as shown in Fig.~\ref{fig-scaling105}. 
In Appendix~\ref{sec:D1problem}, we reconstruct noncollapsing \((D_1,\beta_1)\) trajectories from the single-parameter scaling function by restoring the critical size dependence and converting the scaling variable back to \(N\).

\subsection{Small moments}
\label{sec:beta_smallq}

{Here we generalize the previous $\beta$-function description of the multifractal information-dimension flow to the small-moment regime $D_q(N,W)$ with $q<1/2$, focusing in particular on $q=0.35$. Our aim is to relate its properties to the finite-size scaling approach for these small moments discussed in Section~\ref{sec:moments}, accounting in particular for the strong multifractality controlled by the typical localization length $\xi_\perp$.
}

 {From the finite-size scaling and multifractal analysis presented in Section~\ref{sec:moments}, two regimes within the localized phase $W>W_c$ can be identified, separated by $W^*(q)$. In the interval $W_c<W<W^*$, low-$q$ moments with $q<q^*(W)$ retain a multifractal dependence, with
$
D_q(N,W) \to D_q^\infty(W) = (q/q^*(W)-1)/(q-1)
$
as $N\to\infty$, see Eq.~\eqref{eqstrongMF}. This is a manifestation of bulk exponential localization with $\xi_\perp=q^*(W)/\ln K$ along an exponentially large number of branches. In accordance with the linear scaling of the moments as a function of $\ln N/\xi(W)$, see Eq.~\eqref{eq:Fdeq}, we fit
\begin{equation}
\label{eq:Dq_smallq_multifractal}
D_q(x,W)=D_q^{\infty}(W)+\frac{a(W)}{x}+\frac{b(W)}{x^2},
\qquad 0<D_q^{\infty}<1\;,
\end{equation}
with $x=\log_2 N$.}
This form corresponds to
$P_q\sim N^{-(q-1)D_q^{\infty}}x^{-(q-1)a\ln 2}$, up to the displayed subleading
correction.  It gives
\begin{equation}
\label{eq:beta_smallq_multifractal}
\beta_q(x,W)=
\frac{-a/x^2-2b/x^3}
{\ln2\,[D_q^{\infty}+a/x+b/x^2]}
\longrightarrow0.
\end{equation}
  We also tested {irrelevant power-law corrections in $N$ (data not shown), but the logarithmic expansion Eq.~\eqref{eq:Dq_smallq_multifractal} provided the best fit, consistent with the linear rather than volumic scaling observed for the moments, see Eq.~\eqref{eq:Fdeq}.}

{For the delocalized phase $W<W_c$, we again use a second-order Pad\'e approximant constrained to
$D_q\to 1$. }

\begin{figure}[t]
    \centering
    \includegraphics[width=0.95\linewidth]{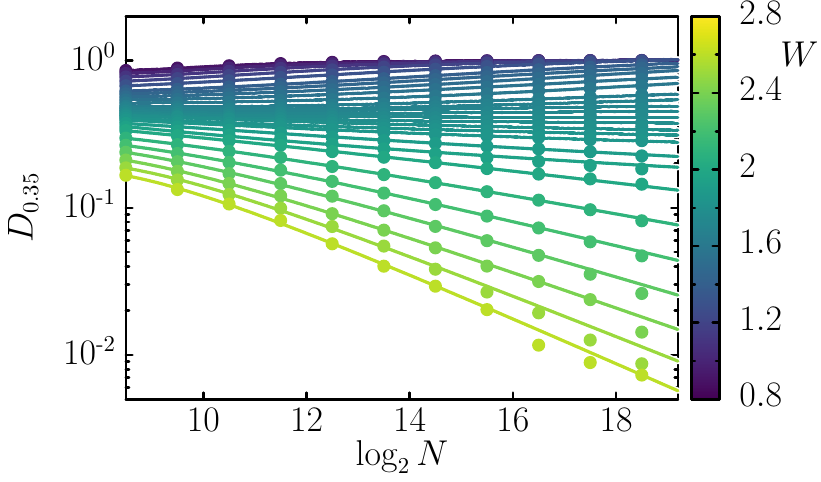}
    \caption{Smoothed multifractal dimension $D_{0.35}$ for $q=0.35<1/2$, for the SWRG with $p=0.06$ as a function of $x=\log_2N$ (symbols) and
    the {localized/multifractal/delocalized} phase-dependent fits (solid curves):  Pad\'e fits are restricted to
    $W<W_c$; logarithmic corrections to a finite $D_q^{\infty}$ are used for
    $W_c<W<W^*$, see Eq.~\eqref{eq:Dq_smallq_multifractal}; the marginal form is used at $W^*$, Eq.~\eqref{eq:Dq_smallq_marginal}; and corrections to a
    localized $P_q$ plateau are used for $W>W^*$, Eq.~\eqref{eq:Dq_smallq_localized}. The loss of relative
    precision at large $N$ and strong disorder follows from differentiating a
    moment that is already close to its limiting constant. 
    }
    \label{fig:D035_fits}
\end{figure}

\begin{figure}[t]
    \centering
\includegraphics[width=0.95\linewidth]{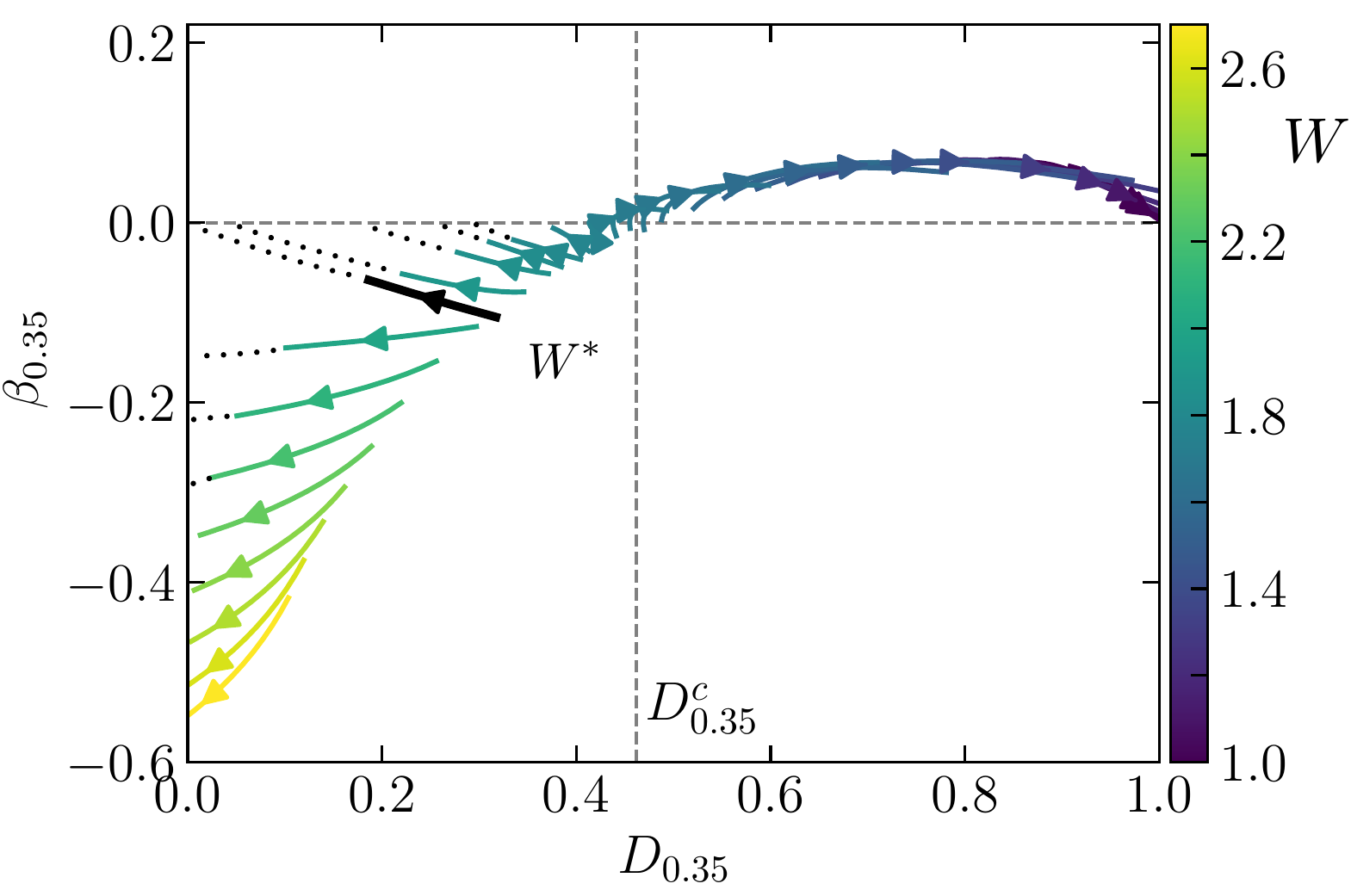}
    \caption{Flow {diagram} $\beta_{0.35}(D_{0.35})$ obtained  from
    the fitted curves of Fig.~\ref{fig:D035_fits} for $q=0.35$. This flow diagram illustrates clearly the strong-multifractal property of the localized phase, with a strong distinction with the flow diagram $\beta_1(D_1)$ shown in Fig.~\ref{fig-beta}. In the localized phase with disorder strength $W_c<W<W^*$, $D_q$ tends at large system sizes to a $D_q^{\infty}>0$ because of the strong-multifractal property manifest in the low-$q$ moments, see text and Eq.~\eqref{eqstrongMF}.  Above $W^*$, the
    moment approaches a localized plateau, $D_q\to0$, and
    $\beta_q\to-\omega(W)$. The dotted lines are eye-guiding continuation of the flow.}
    \label{fig:figbetasmallq}
\end{figure}

At $W=W^*$, for $q=0.35$ we have $W^*\approx 2$,
the asymptotic multifractal dimension $D_q^\infty(W) = (q/q^*-1)/(q-1)$ vanishes. We impose
the marginal form
\begin{equation}
\label{eq:Dq_smallq_marginal}
D_q(x,W^*)=C x^{-1}\exp\left[b/x+c/x^2\right],
\end{equation}
{with again $x=\log_2 N$.}
Hence both $D_q$ and $\beta_q$ approach zero at this point.

For $W>W^*>W_c$, the moment approaches a plateau $P_q \to P_q^{\infty}(W)$ and \(D_q(N,W)\to0\). 
As in the information-dimension case $q= 1>q^*_c=1/2$, see Eq.~\eqref{eq:Pq_rare_branch_correction}, the
leading in $N$ corrections can be written as
\begin{equation}
\label{eq:Pq_smallq_localized}
P_q(N,W)=P_q^{\infty}(W)-A(W)N^{-\omega(W)}
+O\!\left(N^{-2\omega(W)}\right).
\end{equation}
Differentiating this asymptotic expansion leads to
\begin{equation}
\label{eq:Dq_smallq_localized}
D_q(N,W)=C(W)N^{-\omega(W)}
\left[1+a(W)N^{-\omega(W)}\right],
\end{equation}
and therefore
\begin{equation}
\label{eq:beta_smallq_localized}
\beta_q(N,W)=-\omega-
\frac{\omega aN^{-\omega}}{1+aN^{-\omega}}
\longrightarrow-\omega(W)\;,
\end{equation}
as in Eq.~\eqref{eq:beta1_localized_fit}. 

The window-smoothed data for $D_{0.35}$ and the corresponding fits are shown in
Fig.~\ref{fig:D035_fits}, and the flow obtained for $\beta_q$ is shown in
Fig.~\ref{fig:figbetasmallq}.  Deep in the localized phase, the absolute value
of $D_q$ becomes extremely small and its finite-difference estimate is noisy;
the fitted curves are consequently guided primarily by the better sampled
sizes.  
Although it is difficult to distinguish the two regimes, with $D_q^\infty>0$ for $W_c<W<W^*(q)$ and $D_q^\infty=0$ for $W_c<W^*(q)<W$ in the localized phase, the different finite-size behaviors Eq.~\eqref{eq:Dq_smallq_multifractal} for $W<W^*(q)$ and Eq.~\eqref{eq:Dq_smallq_localized} 
derived from our previous FSS analysis, clearly fit the data in the relevant regimes (Fig.~\ref{fig:D035_fits})}.

The beta-function diagrams provide an {interesting} RG representation of the
finite-size behavior contained in the moments $P_q(N,W)$, see Fig.~\ref{fig-scalingsmallq}.
In particular, the asymptotic forms used to fit $D_q$ are chosen
consistently with the moment scaling established in
Sec.~\ref{sec:moments}; the resulting flows provide a very clear distinction between the multifractal regime $D_q^\infty>0$ for $W_c<W<W^*(q)$ and the regime with vanishing asymptotic \(D_q\) for $W_c<W^*(q)<W$.  

\section{A {single}-parameter $\beta$ function for the moments}

In the logic of the conventional formulation of the scaling theory of
localization, we can introduce a
dimensionless running observable and define its beta function as its
logarithmic derivative with respect to the system size
\cite{mackinnon1983scaling,asada2004numerical}. 
Here we can define
\begin{equation}
\beta_X=\frac{d X_q}{d \ln M}
\label{eq:beta_ratio_def}
\end{equation}
where
$X_q\stackrel{\rm def}{=}\ln (P_q/P_q^c)$
 plays the role of the dimensionless observable, and 
 $M=\ln N$ or $M=N$ depending on the type of scaling, linear or volumic, respectively. 

Single-parameter
scaling then corresponds to the statement that this derivative is a
function of the running observable itself, and not independently of the
microscopic parameters or the system size. In the present case this
would amount to
\begin{equation}
    \beta_X(N,W)=f(X_q),
    \label{eq:one_parameter_beta}
\end{equation}
at least within the regime where corrections to scaling can be
neglected \cite{slevin1999corrections}. 

The normalization by the critical moment is crucial for this purpose \cite{maccari2018bkt}.
At the transition,
\begin{equation}
    P_q=P_q^c \ \Rightarrow \     X_q(N,W_c)=0,  \quad
    \beta_X(N,W_c)=0.
    \label{eq:critical_fixed_point}
\end{equation}
for every size.
The critical trajectory is thus mapped onto a fixed point at the origin
of the $(X_q,\beta_X)$ plane, independently of the nontrivial size
dependence of $P_q^c$. {Even in finite dimensions, where $P_q^c \sim N^{-\tau_q}$, normalization by the critical moment has also been used to recover the single-parameter scaling property of the moments across the transition, see, e.g., Ref.~\cite{PhysRevLett.105.046403}. In the present case of small-world graphs, this normalization is particularly important because the critical moments have qualitatively different system-size dependence in different ranges of $q$ \cite{critical2022PRB}.
}

The connection with the finite-size scaling analysis is immediate.
Suppose that, on a branch governed by linear scaling, the normalized
moment satisfies $P_q/P_q^c=F_q(u)$. Then
\begin{equation}
    X_q(N,W)=\ln F_q(u),
    \qquad u=\frac{L}{\xi(W)}.
    \label{eq:X_scaling}
\end{equation}
At fixed disorder,
\begin{equation}
    \frac{du}{d\ln L}=u,
\end{equation}
and Eq.~\eqref{eq:beta_ratio_def} gives
\begin{equation}
    \beta_X=u\,\frac{F_q'(u)}{F_q(u)}.
    \label{eq:beta_parametric}
\end{equation}
If $F_q$ is locally invertible on the branch under consideration,
$u=F_q^{-1}(e^{X_q})$. Eliminating $u$ therefore yields
\begin{equation}
    \beta_X=f(X_q).
    \label{eq:beta_single_parameter}
\end{equation}
Thus single-parameter finite-size scaling implies a single-parameter
beta function on this branch.

{Nevertheless, this $\beta$-function approach provides a stringent and independent test of single-parameter scaling by evaluating Eq.~\eqref{eq:beta_ratio_def} directly from the original moment data, without making use of the fitted scaling functions shown in Figs.~\ref{fig-scaling105} and \ref{fig-scalingsmallq}. We compute}
\begin{equation}
    X_q(N,W)
    =
    \ln P_q(N,W)-\ln P_q^c(N)
\end{equation}
for each disorder and estimate its logarithmic derivative with respect
to $L$ or $N$ from consecutive system sizes. The resulting values of
$\beta_X$ can then be represented directly as a function of $X_q$.

\begin{figure}[t]
    \centering
    \includegraphics[width=0.95\linewidth]{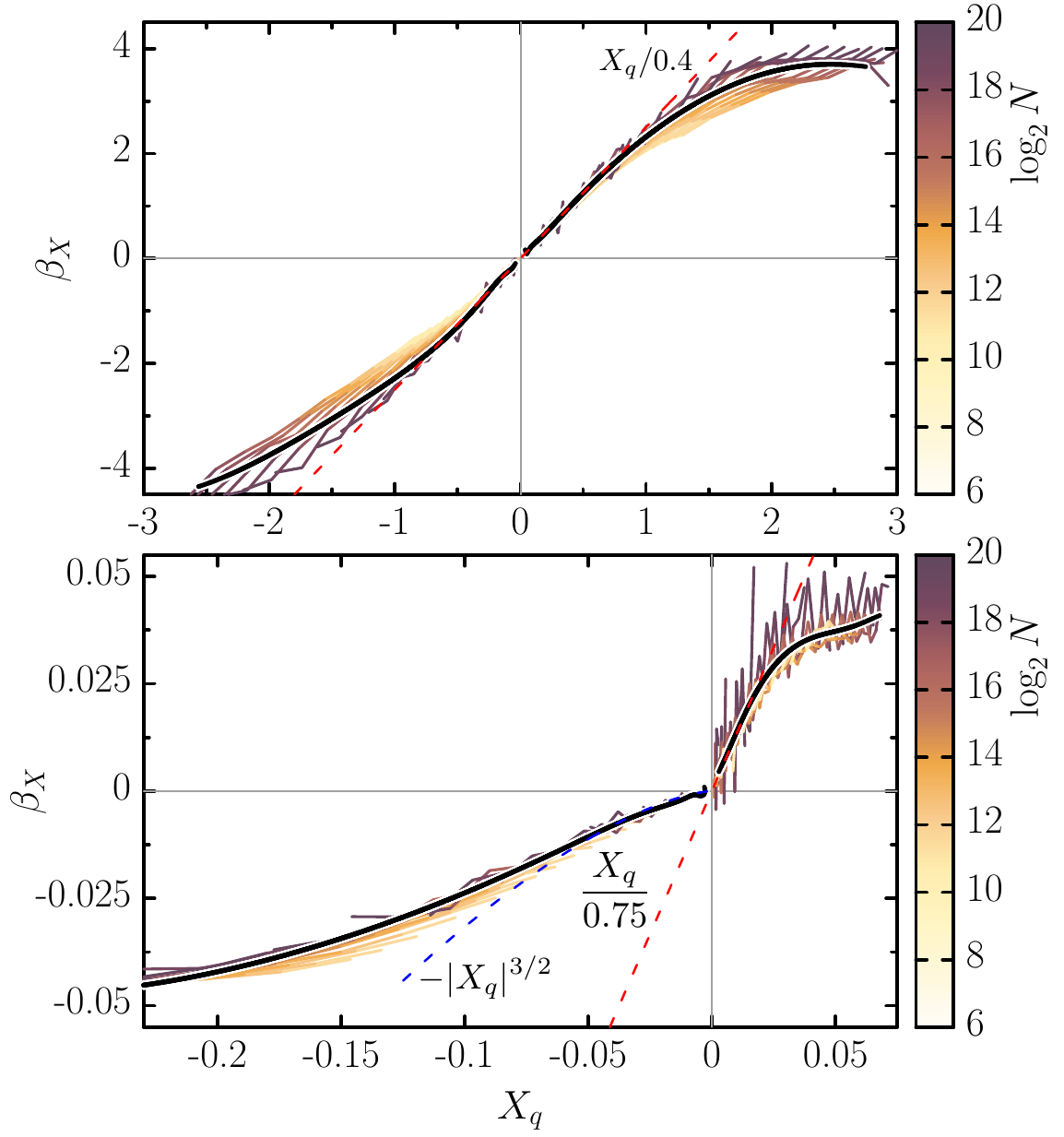} 
    \caption{{Plots of $\beta_X$ with $X_q \stackrel{\rm def}{=} \ln (P_q/P_q^c)$ for $q=0.35$ (top) and $q=1.05$ (bottom). The curves of different colors correspond to data for different system sizes, with the color encoding $\log_2 N$. The black thick line corresponds to $\beta_X$ obtained from the finite size scaling function. The critical point is $(0,0)$. For $q=0.35$ (upper panel) both scalings in each localized and delocalized phases are linear, and the slope at the origin is equal to the inverse of the critical exponent obtained by finite size scaling ($\nu\approx 0.4$, see the red dashed line). For $q=1.05$ (lower panel), this is the case only for the localized phase (right branch), where $\nu_{\rm loc}\approx 0.75$ . In the delocalized phase instead, the scaling is volumic, see Eq.~\eqref{eq:beta_ratio_def} with $|\beta_X(X)|
\propto
|X|^{3/2}$, close to the origin, as indicated by the blue dashed line. This behavior is equivalent to the exponential divergence of the correlation volume Eq.~\eqref{eq:Lambdadiv} found in finite-size scaling, see Fig.~\ref{fig-scaling105}. No window averaging or any
other kind of smoothing of the data was performed. The
only averaging is over random realizations, to obtain $P_q$
and $P_q^c$.  }}
    \label{fig:ratios}
\end{figure}

Remarkably, the data describing the flow with system size in the $(\beta_X,X_q)$ plane for different values of $W$, shown in Fig.~\ref{fig:ratios} for $q=0.35<1/2$ (top panel) and $q=1.05$ (bottom panel), collapse, to a very good approximation, onto a single scaling curve. The color code we use corresponds to the linear size $\log_2 N$, highlighting the fact that the agreement improves with increasing system size. It is important to point out that no window averaging or any other kind of smoothing of the data was performed. The only averaging is over random realizations, to obtain $P_q$ and $P_q^c$. In the $q=1.05$ case, bottom panel of Fig.~\ref{fig:ratios}, the delocalized phase has volumic scaling, so $M=N$, otherwise $M=\ln N$. 
Moreover, this directly reconstructed flow closely follows the beta function obtained by differentiating the finite-size scaling curves in Figs.~\ref{fig-scaling105} and \ref{fig-scalingsmallq}, as shown by the black line.
We stress that the single-parameter scaling demonstrated by this normalized beta function does not necessarily imply a one-dimensional underlying RG flow. The normalization by the critical behavior removes the nontrivial finite-size dependence along the critical trajectory, and the resulting beta function characterizes the scaling of this particular observable \cite{maccari2018bkt}. It therefore provides a complementary perspective to the two-parameter RG description based on \(D_1\), rather than an alternative that necessarily excludes it.

In the localized branch we can make the linear approximation 
\begin{equation}
\beta_X\approx s_q X_q
\end{equation}
where the slope is related to the critical exponent of the transition by $s=1/\nu$ \cite{gangoffour}. The dashed red line in both panels is not obtained from a fit. We purposefully set $s_{q}=1/\nu_{\rm loc}$, obtained from the finite size scaling ($\nu_{\rm loc, q=0.35}\approx 0.4$ and $\nu_{\rm loc, q=1.05}\approx 0.75$). It can be observed that critical exponents obtained from finite size scaling agree well with the slope obtained for $\beta_X$ at the origin.    

{The observed behavior in the delocalized phase for large $q$ requires some further analysis. For the delocalized branch, where the appropriate finite-size scaling is volumic, we have
\begin{equation}
    \beta_X = \frac{dX}{d\ln N}.
\end{equation}
The behavior of $\beta_X$ close to the origin can be related directly to the divergence of the correlation volume,
\begin{equation}
\Lambda \sim \exp\left[A_2(W_c-W)^{-1/2}\right].
\end{equation}
If the beta function were linear near the origin, as is the case for the localized branch (see above), this would lead to a power-law divergence of $\Lambda$, with a critical exponent $\nu_\text{deloc}$ given by the inverse of the slope of the beta function. This is incompatible with the exponential divergence of $\Lambda$ found above. Therefore, the slope of the beta function must vanish at the critical point, $\lim_{X_q\to0^-}\beta_X'(X_q)=0$, as is approximately observed in Fig.~\ref{fig:ratios} (bottom panel).}

{Let us instead assume that the leading behavior is
\begin{equation}
|\beta_X| \propto |X|^p.
\end{equation}
An initial distance from the critical point $|X|\propto W_c-W$ then corresponds to a scale
\begin{equation}
\ln\Lambda
\propto
(W_c-W)^{-(p-1)}.
\end{equation}
Comparing with $\ln\Lambda\propto(W_c-W)^{-1/2}$ gives $p-1=1/2$, or $p=3/2$. We thus expect, close to the critical point,
\begin{equation}
|\beta_X(X)|
\propto
|X|^{3/2},
\end{equation}
and consequently $\lim_{X_q\to0^-}\beta_X'(X_q)=0$.}

{This behavior is consistent with the beta function reconstructed directly from the $P_q$ data: close to the origin, the numerical data are well described by a $3/2$-power dependence, as shown by the blue dashed curve in the bottom panel of Fig.~\ref{fig:ratios}.}

\section{Discussion and conclusion}
\label{sec:discussion}
We have compared two approaches to describing the Anderson transition on small-world random graphs: finite-size scaling of eigenfunction moments and renormalization-group flows of multifractal dimensions. The first approach was developed recently \cite{garciamata2017PRL,twocrit2020PRR,critical2022PRB} to address several important questions concerning the Anderson transition on random graphs. It established the ergodic nature of the delocalized phase and identified two distinct critical localization lengths, leading to a Kosterlitz--Thouless-type flow. The second approach \cite{vanoni2024pnas,altshuler2024renormal} aims to generalize the scaling theory of Anderson localization of the ``gang of four'' \cite{gangoffour} by using the multifractal information dimension in place of the dimensionless conductance. Within this framework, the results indicate asymptotic single-parameter scaling in the delocalized phase, leading to ergodicity, but a two-parameter Kosterlitz--Thouless-type flow in the localized phase.

Although these two descriptions lead to some similar conclusions, the underlying interpretations are different. In the finite-size scaling approach, the two distinct critical length scales are revealed by different sectors of the eigenfunction distribution. In particular, large-order moments, $q>1/2$, follow a single-parameter scaling function controlled by the localization length $\xi_\parallel$, which diverges at the transition with a critical exponent $\nu_\text{loc}\simeq1$. By contrast, small-order moments, $q<1/2$, probe the strong-multifractal structure of the localized phase and are controlled by the distinct typical localization length $\xi_\perp$, associated with the exponent $1/2$. In the renormalization-group approach \cite{vanoni2024pnas,altshuler2024renormal}, by contrast, a single observable, the information dimension, appears sufficient to reveal a two-parameter flow in the localized phase, through RG trajectories that do not collapse onto a single curve.

{Our analysis shows that the RG flows obtained directly from the moments in small-world graphs recover, in the information-dimension sector, the qualitative structure previously found for random regular graphs \cite{vanoni2024pnas,altshuler2024renormal}. Extending the analysis to other moment orders, however, reveals that this flow structure is not unique: its topology depends on which part of the eigenfunction distribution is probed. In particular, the flows for large $q>1/2$ and small $q<1/2$ exhibit qualitatively different behavior.}

{This moment dependence is particularly important in the localized phase. Small moments reveal the strong-multifractal character of the localized phase, governed by the typical localization length $\xi_\perp$. Their flow distinguishes two moment-scaling regimes within the localized phase,
with trajectories approaching either a finite multifractal dimension or zero. The $D_1$ flow therefore constitutes only one sector of a broader, moment-dependent RG description of localization on random graphs.}

{The comparison between the two approaches also highlights the role of the critical normalization used in finite-size scaling. Normalizing the moments by their critical values, $P_q/P_q^c$, removes the non-trivial finite-size dependence of the critical trajectory and allows one to focus on the evolution away from criticality towards the localized or delocalized phases. The resulting flow obeys single-parameter scaling. By contrast, the flow of the multifractal dimensions retains the non-trivial system-size dependence of the critical moments. The resulting family of RG trajectories can therefore reflect the finite-size structure of the critical point without necessarily implying the presence of a second independent scaling variable.}

{We have explicitly constructed a $\beta$-function description for the normalized moments $P_q/P_q^c$. Computing the beta function directly from the raw moment data, without any smoothing or fitting procedure, yields RG trajectories that collapse, to a very good approximation, onto a common scaling curve over the available system sizes. This provides an independent and stringent test of single-parameter scaling. At the same time, the moment beta function retains information about the critical behavior. On the localized branches, its behavior near the fixed point is consistent with the two localization-length critical exponents obtained independently from finite-size scaling. On the volumic delocalized branch, its vanishing slope and nonlinear behavior are consistent with the exponential divergence of the correlation volume.}

The critical exponents extracted here are consistent with the exponents $1/2$ and $1$ that also enter the RG descriptions \cite{critical2022PRB,vanoni2024pnas,altshuler2024renormal}, although their physical assignments are not identical. The additional delocalized scale with exponent one proposed in one asymptotic scenario is not resolved at the system sizes accessible here. Establishing whether this scale emerges at larger volumes, and determining how corrections to scaling affect the different moment sectors, remain open questions.

Our results therefore highlight an interesting tension between two complementary scaling descriptions. On the one hand, the RG approach of Refs. \cite{vanoni2024pnas,altshuler2024renormal,niedda2024renormal} identifies a two-dimensional flow structure from the evolution of a single observable $D_1$. On the other hand, our finite-size scaling analysis reveals two distinct scaling sectors by considering eigenfunction moments with $q>1/2$ and $q<1/2$, each of which separately exhibits single-parameter scaling. The beta function constructed from moments normalized by their critical behavior 
provides a direct check of this scaling using the moment data, while the beta function of the multifractal dimensions retains the nontrivial finite-size evolution of the critical moments. These observations suggest that the collapse or noncollapse of the beta function associated with a single observable does not, by itself, determine the dimensionality of the underlying RG flow. Our results thus emphasize the complementary nature of the two approaches and the importance of considering different observables to obtain a more complete picture of the scaling structure of the Anderson transition on random graphs.


\appendix
\section{Obtaining the $q=1$ flow from the single-parameter scaling function}
\label{sec:D1problem}
{In this appendix, we derive the $q=1$ beta function $\beta_1(N,W)$ for the multifractal information dimension $D_1$ defined in Sec.~\ref{sec:beta} from the single-parameter scaling function of the normalized moment $P_q/P_q^c$ described in Section \ref{sec:moments}. The purpose is to clarify how a single-parameter scaling function can generate a family of trajectories in the flow diagram on the $(D_q,\beta_q)$ plane.}

{As shown in Section \ref{sec:moments}, the moments obey a single-parameter scaling property that can be written as}:
\begin{equation}
\label{eq:Fq_scaling_sectionV}
F_q(N,W)=\frac{P_q(N,W)}{P_q^c(N)}=\mathcal F_q(X),
\end{equation}
where $X=N/\Lambda(W)$ for volumic scaling and
$X=\ln N/\xi(W)$ for linear scaling. 

{Let us define the contribution associated}
with the scaling function by
\begin{equation}
\label{eq:normalized_DF}
D_q^F(X)=-\frac{1}{q-1}\frac{d\ln\mathcal F_q(X)}{d\ln X}
\end{equation}
and
\begin{equation}
\label{eq:normalized_betaF}
\beta_q^F(X)=\frac{d\ln|D_q^F(X)|}{d\ln X}.
\end{equation}
The absolute value is needed because the contribution of the normalized
moment can have different signs on the two sides of the transition.  Since
both quantities in Eqs.~\eqref{eq:normalized_DF} and
\eqref{eq:normalized_betaF} are obtained from the same function
$\mathcal F_q(X)$, the parametric representation
$\beta_q^F(D_q^F)$ is single-parameter by construction.

The scaling function-derived flow is not, however, the same as the one defined from $D_q$.  To connect
the two, let us {introduce the scale conversion contribution:
\begin{equation}
\kappa(X,W)=\frac{d\ln X}{d\ln N},
\end{equation}
and the critical contribution:}
\begin{equation}
\label{eq:Dqc_sectionV}
D_q^c(N)=-\frac{1}{q-1}\frac{d\ln P_q^c(N)}{d\ln N},
\qquad
\beta_q^c(N)=\frac{d\ln D_q^c(N)}{d\ln N}.
\end{equation}

Differentiating Eq.~\eqref{eq:Fq_scaling_sectionV} gives
\begin{equation}
\label{eq:Dq_from_normalizedF}
D_q(N,W)=D_q^c(N)+\kappa(X,W)D_q^F(X).
\end{equation}
Taking one more derivative, and defining
$\gamma=d\ln\kappa/d\ln N$, yields
\begin{equation}
\label{eq:beta_from_normalizedF}
\beta_qD_q
=\beta_q^cD_q^c
+\kappa D_q^F\left(\gamma+\kappa\beta_q^F\right).
\end{equation}
Equations~\eqref{eq:Dq_from_normalizedF} and
\eqref{eq:beta_from_normalizedF} are the required transformations from the
single-parameter flow of $P_q/P_q^c$ to the flow of multifractal dimension $D_q$.
They {clearly} show that single-parameter scaling of $F_q$ does not necessarily imply
a single curve in the $(D_q,\beta_q)$ plane:  The $D_q$ flow also contains
the critical trajectory and the conversion between the scaling coordinate
$X$ and the system size $N$.

{Let us examine more closely the contribution from the scale conversion. We distinguish two cases, corresponding to volumic and linear scaling.}

{For volumic scaling, $X=N/\Lambda(W)$, and therefore
\begin{equation}
\kappa=1,
\qquad
\gamma=0.
\end{equation}
The transformation then reduces to
\begin{equation}
\label{eq:volumic_reconstruction}
D_q=D_q^c+D_q^F,
\qquad
\beta_q D_q=\beta_q^c D_q^c+\beta_q^F D_q^F.
\end{equation}
The correlation volume only translates the logarithmic size coordinate,
\begin{equation}
\ln N=\ln X+\ln\Lambda(W).
\end{equation}}

{For linear scaling, $X=\ln N/\xi(W)$. In this case,
\begin{equation}
\kappa=\frac{1}{\ln N},
\qquad
\gamma=-\frac{1}{\ln N},
\end{equation}
and Eqs.~\eqref{eq:Dq_from_normalizedF} and
\eqref{eq:beta_from_normalizedF} become
\begin{equation}
\label{eq:linear_D_reconstruction}
D_q=D_q^c+\frac{D_q^F}{\ln N},
\end{equation}
and
\begin{equation}
\label{eq:linear_beta_reconstruction}
\beta_q D_q
=\beta_q^c D_q^c
+\frac{D_q^F}{(\ln N)^2}\left(\beta_q^F-1\right).
\end{equation}
Since $\ln N=\xi(W)X$, the factors appearing in
Eqs.~\eqref{eq:linear_D_reconstruction} and
\eqref{eq:linear_beta_reconstruction} retain an explicit dependence on the
correlation length. Linear scaling therefore generally produces a family of flow curves. In geometrical terms, volumic scaling translates the $\ln N$
coordinate, whereas linear scaling rescales it; the latter changes the rate at
which the flow trajectory is traversed.}

{Let us now examine the critical contribution. If $P_q^c$ is a pure power of $N$, then $D_q^c$ is constant and $\beta_q^c=0$, and hence there is no contribution from the critical term. If the critical moment instead contains a logarithmic size dependence, $D_q^c$ and $\beta_q^c$ continue to depend on the absolute system size $N$. At a given value of $X$, different values of $W$ then sample different points along the critical trajectory because of the shift $\ln\Lambda(W)$. This contributes to the emergence of a family of trajectories.}

{As we showed in Section~\ref{sec:moments}, volumic scaling occurs for $q=1$ on the delocalized side, which is associated with a trivial scale-conversion contribution. The critical behavior, however, is of logarithmic multifractality type,
$
P_q^c\sim(\ln N)^{-d_q(q-1)}
$
\cite{PhysRevResearch.6.L032024,PhysRevB.110.014210}. The critical behavior therefore corresponds to a curve rather than a fixed point in the $(D_q,\beta_q)$ plane. Combining this running critical contribution with the single-parameter flow $\beta_q^F$ through Eq.~\eqref{eq:volumic_reconstruction} produces a family of size- and disorder-dependent trajectories. Nevertheless, since the critical behavior depends only logarithmically on $N$, its contribution becomes negligible at large system sizes compared with the volumic scaling flow, and we recover the asymptotic single-parameter scaling observed in \cite{vanoni2024pnas,altshuler2024renormal,niedda2024renormal}. }

{In the localized phase for $q=1$, we have shown instead that the FSS is linear. The relations above therefore show explicitly that the reconstructed physical trajectory retains the factor $\xi(W)$ through $\ln N=\xi(W)X$. Thus, even though $\mathcal F_q(X)$ defines a single-parameter flow, the conversion back to $(D_1,\beta_1)$ can produce a family of disorder-dependent trajectories.}

\begin{figure}[ht]
    \centering
    \includegraphics[width=0.95\linewidth]{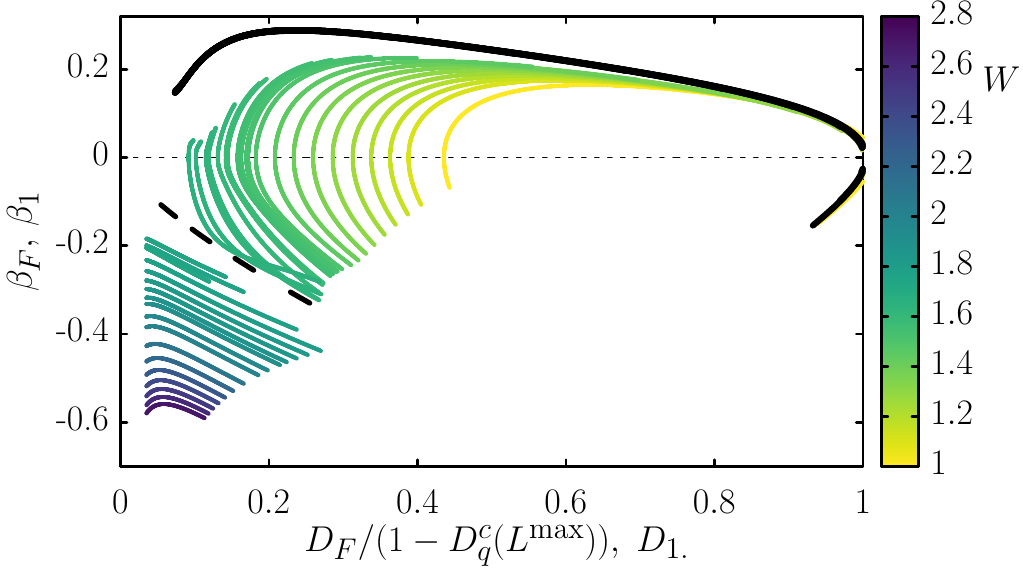} 
\caption{Flow reconstructed for $q=1$ from the volumic
    single-parameter scaling function of the normalized moments and the fitted
    critical moments. The thick black line shows $\beta_q^F)$ as a function of the rescaled variable $F_F/(1-D_q^c(L^{\rm max}))$ determined by    the single variable $X=N/\Lambda(W)$ in the delocalized  phase.  The colored trajectories show the
    physical $(D_q,\beta_q)$ flow obtained after adding the running critical
    contribution through Eq.~\eqref{eq:volumic_reconstruction}. The critical flow is represented by the dashed-black line}
    \label{fig:fitbetaF}
\end{figure}

{We represent in Fig.~\ref{fig:fitbetaF} the flow reconstructed from the single-parameter FSS scaling function. The procedure is as follows. We fitted the collapsed values of $\ln F_q$ (see Fig.~\ref{fig-scaling105} for $q=1.05$) as a smooth function of the logarithm of the appropriate scaling coordinate, using a smoothing regression rather than interpolating the individual data points. The critical moment was fitted separately. For each disorder strength, the appropriate correlation length or correlation volume obtained from the FSS analysis was then used to evaluate the fitted scaling function on a dense size grid and reconstruct
$P_q(N,W)=P_q^c(N)F_q(N,W).$
The reconstructed moments were differentiated (logarithmic) numerically with respect to the system size $N$ to obtain $D_q$, and their logarithmic derivative was then taken to obtain $\beta_q$, following the same definitions used for the original moment data. Only points within the range covered by the corresponding FSS data were retained in the flow curves. In Fig.~\ref{fig:fitbetaF}, we show the flow for $q=1$, obtained by reconstructing $P_{0.95}$ and $P_{1.05}$ as described above, and then computing $D_1$ using the symmetric construction of Eq.~\eqref{eq:D1_symmetric}.}

This reconstruction therefore establishes a direct relation between the two descriptions used in this work. The moments possess a single-parameter scaling property. The beta function for the multifractal dimension $D_q$, however, is obtained only after restoring the critical behavior size-dependence and converting the scaling coordinate back to $N$. A running critical trajectory or the Jacobian associated with linear scaling can then transform a single-parameter scaling function into 
a family of disorder-dependent trajectories
observed in the $(\beta_1,D_1)$ flow diagram.


\begin{acknowledgments}
The authors acknowledge support from  Argentinian Agency ANCyPT (Grant No. PICT-2020-SERIEA-00740), from the EUR grant NanoX ANR-17-EURE-0009 in the framework of the “Programme des Investissements d’Avenir”,
the France 2030 program (Grant No. ANR-23-PETQ-0002), {from research funding grant ManyBodyNet No. ANR-24-CE30-5851, and from Singapore Ministry of Education Academic Research Funds Tier II (MOE-T2EP50223-0009
and MOE-T2EP50222-0005).} In addition, support from CNRS (France) through the International Research Project (IRP) ``Complex Quantum Systems'' (CoQSys) is gratefully acknowledged. Computational resources were provided by Calcul en Midi-Pyr\'en\'ees (CALMIP) and by the Consortium des Equipements de Calcul Intensif (CECI), funded by the Fonds de la Recherche Scientifique de Belgique (F.R.S.-FNRS) under Grant No. 2.5020.11.

\end{acknowledgments}

\bibliography{refs_merged}

\end{document}